\documentclass[conference]{IEEEtran}
\usepackage{cite}
\usepackage{amsmath,amssymb,amsfonts}
\usepackage{algorithmic}
\usepackage{pgf-pie}
\usepackage{graphicx}
\usepackage{amsmath}
\usepackage{bm}
\usepackage{subcaption}
\usepackage{booktabs}
\usepackage{textcomp}
\usepackage{xcolor}
\usepackage{array}
\usepackage{graphicx}
\usepackage{textcomp}
\usepackage{geometry}

\newcommand{\setpapergeometry}{
  \ifdim\paperwidth=8.5in % US Letter
    \geometry{
      paper=letterpaper,
      top=0.75in,
      bottom=1in,
      left=0.625in,
      right=0.625in
    }
  \else
    \ifdim\paperwidth=210mm % A4
      \geometry{
        paper=a4paper,
        top=19mm,
        bottom=43mm,
        left=13mm,
        right=13mm
      }
    \else
      \geometry{} % Default geometry settings if paper size is unknown
    \fi
  \fi
}

\setpapergeometry % Apply the appropriate margin settings
\usepackage{xcolor}
\usepackage[utf8]{inputenc}
\usepackage[justification=centering, font=footnotesize]{caption}
\usepackage[nonumberlist,nopostdot,nomain,nogroupskip]{glossaries}
\usepackage[nolist]{acronym}
\begin{acronym}
\acro{3DN}{3D Network}
\acro{3GPP}{3rd Generation Partnership Project}
\acro{ATG}{air-to-ground}
\acro{BS}{Base Station}
\acro{CA}{Cooperative Awareness}
\acro{CAM}{Cooperative Awareness Message}
\acro{CAV}{connected and autonomous vehicle}
\acro{CP}{Collective Perception}
\acro{CPS}{Collective Perception Service}
\acro{CPM}{Collective Perception Message}
\acro{CSI}{channel state information}
\acro{DF}{Decode-and-Forward}
\acro{D2D}{device-to-device}
\acro{DFT}{Discrete Fourier Transform}
\acro{E2E}{End-to-End}
\acro{FoV}{field of view}
\acro{FD}{Full-Duplex}
\acro{GoB}{Grid-of-Beams}
\acro{GUE}{Ground User Equipment}
\acro{HBF}{hybrid digital beamforming}
\acro{IoT}{Internet of Things}
\acro{IRS}{Intelligent Reflective Surface}
\acro{ITS}{Intelligent Transport System}
\acro{ITS-S}{Intelligent Transport System Station}
\acro{KPI}{Key Performance Indicator}
\acro{mBS}{Micro Base Station}
\acro{MBS}{Macro Base Station}
\acro{mmWave}{millimiter-wave}
\acro{MU-MIMO}{Multi-User Multiple-Input Multiple-Output }
\acro{LoS}{line of sight}
\acro{NLoS}{Non-\ac{LoS}}
\acro{OFDM}{Orthogonal Frequency-Division Multiplexing}
\acro{QoE}{Quality of Experience}
\acro{QoS}{Quality of Service}
\acro{PL}{path loss}
\acro{PPP}{Poisson Point Process}
\acro{pmf}{probability mass function}
\acro{PRB}{Physical Resource Block}
\acro{RB}{Resource Block}
\acro{RF}{radio-frequency}
\acro{RRM}{Radio Resource Management}
\acro{RRA}{Radio Resources Assignment}
\acro{RSU}{Road Side Unit}
\acro{RU}{Resource Unit}
\acro{SNR}{Signal-to-Noise Ratio}
\acro{TR}{Technical Report}
\acro{TBS}{Terrestrial Base Station}
\acro{ULA}{Uniform Linear Array}
\acro{UAV}{Unmanned Aerial Vehicle}
\acro{UABS}{Unmanned Aerial Base Station}
\acro{UE}{User Equipment}
\acro{UMa}{Urban Macro}
\acro{UMi}{Urban Micro}
\acro{V2V}{Vehicle-to-Vehicle} 
\acro{V2X}{Vehicle-to-Anything}
\acro{5G}{Fifth-Generation}

\end{acronym}
\def\BibTeX{{\rm B\kern-.05em{\sc i\kern-.025em b}\kern-.08em
    T\kern-.1667em\lower.7ex\hbox{E}\kern-.125emX}}
    \usepackage{amsmath,amssymb,amsfonts}

\begin{document}
%\bstctlcite{IEEEexample:BSTcontrol}

\title{A Thorough Analysis of \\ Radio Resource Assignment  for UAV-Enhanced Vehicular Sidelink Communications\vspace{-.3cm}}
\author{\IEEEauthorblockN{
Francesca Conserva\IEEEauthorrefmark{1},
Francesco Linsalata\IEEEauthorrefmark{2},
Marouan Mizmizi\IEEEauthorrefmark{2},
Maurizio Magarini\IEEEauthorrefmark{2},
Umberto Spagnolini\IEEEauthorrefmark{2},\\
Roberto Verdone\IEEEauthorrefmark{1} and
Chiara Buratti\IEEEauthorrefmark{1}}
\IEEEauthorblockA{\IEEEauthorrefmark{1}
\small Dipartimento di Ingegneria dell'Energia Elettrica e dell'Informazione, University of Bologna, \& WiLab, CNIT, \textit{Italy}}
\IEEEauthorblockA{\IEEEauthorrefmark{2}
\small Dipartimento di Elettronica, Informazione e Bioingegneria, Politecnico di Milano, Milan, \textit{Italy}}
\thanks{.}
\vspace{-.6cm}
}
\maketitle

\begin{abstract}
The rapid expansion of \acp{CAV} and the shift towards \ac{mmWave} frequencies offer unprecedented opportunities to enhance road safety and traffic efficiency. Sidelink communication, enabling direct \ac{V2V} communications, play a pivotal role in this transformation. As communication technologies transit to higher frequencies, the associated increase in bandwidth comes at the cost of a severe path and penetration loss. In response to these challenges, we investigate a network configuration that deploys beamforming-capable \acp{UAV} as relay nodes. In this work, we present a comprehensive analytical framework with a groundbreaking performance metric, i.e. average access probability, that quantifies user satisfaction, considering factors across different protocol stack layers. Additionally, we introduce two \ac{RRA} methods tailored for \acp{UAV}. These methods consider parameters such as resource availability, vehicle distribution, and latency requirements. Through our analytical approach, we optimize the average access probability by controlling \ac{UAV} altitude based on traffic density.
Our numerical findings validate the proposed model and strategy, which ensures that \ac{QoS} standards are met in the domain of \ac{V2X} sidelink communications.

\end{abstract}
\begin{IEEEkeywords}
Unmanned Aerial Vehicles, V2X, Radio Resource Assignment, Beamforming, QoS.
% \vspace{-.30cm}
\end{IEEEkeywords}

\section{Introduction}
\label{sec:intro}
\begin{figure*}[t!]
\centering
\includegraphics[width=1\textwidth]{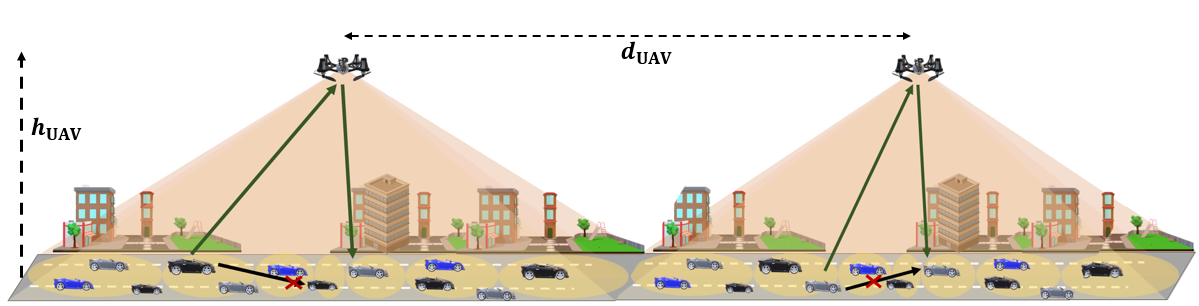}
\caption{UAV-Enhanced vehicular sidelink communications scenario: \acp{CAV}, seeking to establish communication with other \acp{CAV}, but facing a blocked \ac{V2V} link, exploit \ac{UAV} as relay.}
\vspace{-0.5cm}
\label{fig:scenario}
\end{figure*}

The widespread deployment of \acp{CAV} is currently underway, holding the potential to revolutionize global mobility by significantly enhancing road safety and efficiency \cite{sidelink, ITSCost}. Notably, \ac{V2X} communications, with specific emphasis on sidelink technology, facilitate direct \ac{V2V} interaction via the PC5 interface \cite{sidelink2}. This technological advancement serves as a pivotal enabler for a multitude of vehicular applications, including, but not limited to, extended sensing, cooperative awareness, and coordinated maneuvers \cite{5GAA}.
These applications exhibit a considerable demand for data rate, which enforces a shift towards the \ac{mmWave} spectrum ($>$24 GHz) to explore the large available bandwidth \cite{mmWave}. However, communication at \ac{mmWave} frequencies entails using beamforming systems to counteract the severe path and penetration loss associated with the propagation of high-frequency signals. In contexts characterized by high mobility, such as in \ac{V2X} scenarios, \ac{LoS} blockages become increasingly prevalent as traffic density rises, consequently posing a substantial challenge to communication reliability \cite{LinsalataLoSMap}. 
An effective strategy to mitigate this issue is to implement relay-assisted transmissions. Various relay mechanisms have been investigated in the literature, including opportunistic relays, \acp{RSU}, and \acp{IRS} \cite{LinsalataLoSMap, TVT_self_org, TVT_irs}.

Relay-aided communications, particularly leveraging \acp{UAV}, have gained significant attention for enhancing various aspects of communication systems~\cite{FD, SG, novelapp}. In the field of \acp{ITS}, \acp{UAV} can be vital during disasters or network failures. The paper \cite{ITS} proposes a method for predicting information dissemination through sidelink communication to guide \ac{UAV} flight trajectories, swiftly disseminating information to vehicles. In \cite{offloading}, the authors explore the potential of utilizing \acp{UAV} in vehicular networks for task offloading. Their objective is to maximize the number of offloaded tasks by jointly optimizing the initial \ac{UAV} launch point and strategically associating \acp{CAV} to the \ac{UAV}. In \cite{Wu}, \acp{UAV} are also examined, investigating their utilization for supporting edge caching in terrestrial vehicular networks while maximizing the overall network throughput by formulating a joint caching and trajectory optimization problem. In \cite{Khabbaz}, \acp{UAV} serve as relays to mitigate the rapid topology changes impacting vehicular networks, leading to connectivity interruptions and data delivery delays.
The authors of \cite{jointopt} address \ac{UAV}-aided \ac{mmWave} vehicular networks. Their objective is to minimize time consumption while meeting traffic demands through efficient resource allocation schemes.

These works, although relevant in the context of using \ac{UAV} as relays, have predominantly focused on providing analytical frameworks limited to specific domains, encompassing connectivity assessment \cite{Khabbaz}, optimal UAV deployment \cite{ITS}, trajectory optimization\cite{Wu, ITS}, throughput maximization \cite{Wu}, or relay selection \cite{offloading}.

In this article, we introduce an enhanced approach to sidelink communications by adopting beamforming-enabled  \acp{UAV} as relay nodes. This ensures the reliable exchange of data on the sidelink data plane, even in scenarios where \ac{LoS} communication is obstructed. Particularly, building upon \cite{FC}, our intention is to advance the current state of knowledge by introducing a comprehensive analytical framework that features a novel performance metric: the average access probability. The latter numerically quantifies the average user satisfaction following the \ac{RRA} process when more \acp{CAV} are covered by an \ac{UAV} and request resources to fulfill traffic demand.

In contrast with previous work, our mathematical framework concurrently considers aspects that span across multiple layers of the protocol stack. Specifically, the proposed framework accounts for: i) on-board beamforming; ii) coverage analysis, dealing with the statistics of the \ac{SNR}; iii) radio resource assignment strategies; iv) distribution of underlying vehicles, \ac{UAV} altitude, available bandwidth and v) application-dependent \ac{V2X} traffic demand from \acp{CAV}. Notably, none of the aforementioned works concentrate on any particular \ac{V2X} use case while accounting for the high vehicular traffic demand, which is a crucial aspect to manage.

\textbf{Contributions} %In the following, 
The main contributions of this research can be summarized as follows: %are highlighted.

\begin{itemize}
\item An analytical characterization of the average access probability is proposed. This analysis emphasizes the impact of critical system parameters, the \ac{RRA} process, traffic dynamics, and \ac{V2X}  \ac{QoS} requirements.
\item Two distinct \ac{RRA} algorithms are also presented. These algorithms consider factors such as vehicle distribution, beamforming, and available bandwidth, resulting in varying performance outcomes depending on the selected scenario.
\end{itemize}

\textbf{Organization} 
The remainder of the paper is organized as follows. Section II defines the system model. The RRA algorithms and their performances are characterized in Sec. III.
The simulation results are presented in Sec. IV. Lastly, Sec. V concludes the work.

\section{System Model} 
\label{sec:refscen}

Let us examine the \ac{V2V} communication scenario supported by \acp{UAV}, as illustrated in Figure \ref{fig:scenario}. In this setting, $M$ \acp{CAV} are in transit along a highway, while $Q$ \acp{UAV} are positioned above the \acp{CAV} at an altitude $h_{\rm UAV}$. The \acp{UAV} are evenly distributed along the road at a fixed distance of $d_{\rm UAV}$ from one another, and each \ac{UAV} is responsible for covering a specific road segment with a length of $L_{\rm f}$. We assume that each \ac{UAV} has a hybrid digital \ac{ULA} with $N_{\text{UAV}}$ antenna elements and $N_\mathrm{RF}$ \ac{RF} chains, while each \ac{CAV} equips an \ac{ULA} with $N_\mathrm{CAV}$ antenna elements and a single \ac{RF} chain.
When the $i$th \ac{CAV} detects a communication disruption with the $j$th \ac{CAV}, a request for relay assistance is initiated and directed towards the $q$th \ac{UAV} responsible for covering the road segment corresponding to the $i$th \ac{CAV}.
\subsection{UAV Beamforming Design}
The \acp{UAV} utilize a set of beams, denoted as $\mathcal{B}$, which serves as a codebook for both uplink and downlink operations. Similarly to the 5G NR Standard \cite{3gpp}, we assume that each communication instance occurs in a pre-defined time-frequency resource. A scheduling strategy is implemented at the \ac{UAV} to ensure that \acp{CAV} served by the same beam are assigned distinct time-frequency resources to prevent interference, whereas \acp{CAV} covered by different beams can share the same time-frequency resources. The association between beams and \acp{CAV} can be established during the initial access procedure as in \cite{MorandiPCB}.

Since the \ac{UAV} employs a \ac{HBF} architecture with $N_\mathrm{RF}$ chains, for each time-frequency resource, the scheduler allocates $N_\mathrm{RF}$ \acp{CAV} that are simultaneously transmitting towards the \ac{UAV}.
The signal transmitted by the $i$th \ac{CAV} is $\mathbf{x}_i = \mathbf{f}_i s_i$, where $\mathbf{f}_i \in \mathbb{C}^{N_\mathrm{CAV} \times 1}$ denotes the beamforming vector and $s_i$ denotes the transmitted symbol such that $\mathbb{E}\left[s_i^* s_i\right] = 1$. After time synchronization, the discrete-time signal received by the \ac{UAV}, in a generic time-frequency resource, can be expressed as
\vspace{-0.3cm}
\begin{align} \label{eq: received signal}
    \mathbf{y} = \sum_{i=1}^{N_\mathrm{RF}} \sqrt{\rho_i}\,\mathbf{H}_i\,\mathbf{x}_i + \mathbf{n} \,,
\end{align}
where $\mathbf{H}_i \in \mathbb{C}^{N_\mathrm{UAV} \times N_\mathrm{CAV}}$ is the channel impulse response such that $\mathbb{E}\left[\|H_i\|_\mathrm{F}^2\right] = 1$, $\mathbf{n} \sim \mathcal{CN}(0, \sigma_n^2)$ denotes the additive Gaussian noise, and $\rho_i$ represents the average received power from the $i$th \ac{CAV}, defined as
\begin{equation}
    \rho_i = \frac{N_\mathrm{UAV} \, N_\mathrm{CAV} \, P_i}{PL_i},
\end{equation}
where $P_i$ denotes the transmitted power and $PL_i$ is the path loss.
The signal from the $i$th \ac{CAV} decoded at the \ac{UAV} can be expressed as
\begin{equation}
    \hat{s}_i = \sqrt{\rho_i} \mathbf{w}_\ell^\mathrm{H} \, \mathbf{H}_i\,\mathbf{x}_i + \mathbf{w}_\ell^\mathrm{H} \, \mathbf{n},
\end{equation}
where $\mathbf{w}_\ell \in \mathbb{C}^{N_\mathrm{UAV} \times 1}$ is the beamforming vector applied at \ac{UAV}, that is drawn from the codebook $\mathbf{w}_\ell \in \mathcal{B}$. Throughout the paper, we consider a \ac{DFT} codebook as in \cite{DFT}, which ensures full coverage with the minimum number of beams. 
The resulting instantaneous \ac{SNR} is given by
\vspace{-0.3cm}
\begin{equation}\label{eq:SNRdef}
    \mathrm{SNR}_i = \frac{\rho_i \, \|\mathbf{w}_\ell^\mathrm{H} \,\mathbf{H}_i \, \mathbf{f}_i\|^2} {\sigma_n^2}.
\end{equation}

\subsection{Channel Model}
The strategic placement of the \ac{UAV} ensures a reliable ground-to-aerial communication link without obstructions. The path loss in dB experienced by the link from the $i$th \ac{CAV} can be expressed as in \cite{shakhatreh2021modeling}:
\begin{equation}
\label{PL}
PL_{i} = A + \alpha \,10 \log_{10}(d_{i}) + \eta,
\end{equation}
where $d_{i}$ denotes the distance between the \ac{UAV} and the $i$th \ac{CAV}, $A$ and $\alpha$ denote the excess path loss offset and the path loss exponent, respectively, while $\eta \sim \mathcal{N}(0,\sigma_{s}^2)$ is the log-normal shadowing component, with variance $\sigma_s^2$. The multipath channel impulse response matrix is modeled as %in \textcolor{red}{[REF]}
\vspace{-0.3cm}
\begin{align} 
    \mathbf{H} = \sum_{p = 1}^P \beta_p \mathbf{a}_\mathrm{UAV}^\mathrm{H}(\theta_p) \mathbf{a}_\mathrm{CAV}^\mathrm{H}(\phi_p),
\label{eq:channel_matrix_compact}
\end{align}
where $\beta_p \sim \mathcal{CN}(0, \sigma_p^2)$ denotes the complex amplitude of the $p$th path, such that $\sum_p \sigma_p^2 = 1$, $P$ is the number of multipath, $\phi_p$ and $\theta_p$ denote the angle of departure and angle of arrival, respectively, and $\mathbf{a}(\theta)$ is the normalized array function \cite{mizCha}.

\section{Performance assessment of UAV-assisted radio resource assignment}
\label{sec:RRA}

This section highlights the \ac{RRA} procedures and how they impact the average access probability. 
 
\subsection{Radio Resources Assignment Algorithms}

Two \ac{RRA} algorithms are investigated accounting for the specific beamforming technique, the available radio resources at the \acp{UAV}, and the \ac{E2E} application latency requirement, which we denote by $\tau_\mathrm{E2E}$. 

Let us consider a grid of time-frequency resources with $N_\mathrm{ch} \times N_\mathrm{slot}$, where $N_\mathrm{ch}$ denotes the number of frequency subchannels and $N_\mathrm{slot}$ is the number of temporal slots. A similar architecture is used by the 5G NR Standard \cite{3gpp}, where $N_\mathrm{ch}$ depends on the allocated bandwidth and $N_\mathrm{slot} \leq (\tau_\mathrm{E2E} \, /T_\mathrm{slot}) \, 0.5$ is defined according to the \ac{E2E} latency requirement, with $T_\mathrm{slot}$ being the slot duration. Note that the product with $0.5$ accounts for the resources to forward the data from the \ac{UAV} to the receiving \acp{CAV}.

\subsubsection{Fair RRA}

The first approach, denoted as Fair \ac{RRA}, ensures an even resource allocation between the set of beams and can be considered as a baseline. The number of time-frequency resources for the $\ell$th beam is
\vspace{-0.2cm}
\begin{equation}\label{fair}
    N^{(\text{Fair})}_\ell= \left\lfloor  \frac{N_{\text{slot}} N_{\text{ch}} N_{\text{RF}}}{N_{\text{beam}}} \right\rfloor ,
\end{equation}
where $N_\mathrm{beam} = |\mathcal{B}|$ is the cardinality of the \ac{UAV} beamforming codebook.

\subsubsection{Beam-based RRA}

The second \ac{RRA} approach considered, denoted as beam-based (BB) \ac{RRA}, entails the distribution of resources among the various beams, taking into account the size of each specific beam footprint. Indeed, the larger the beam footprint, the higher the number of \acp{CAV} that will be covered and should be served. Accordingly, the number of time-frequency resources allocated for the $\ell$th beam is
\begin{equation}
\label{beambased}
    N^{(\text{BB})}_\ell = \left\lfloor N_{\text{slot}} N_{\text{ch}} N_{\text{RF}} \frac{L_\ell}{L_{\rm f}} \right\rfloor,
\end{equation}
where $L_\ell$ denotes the $\ell$th beam footprint such that $L_{\rm f} = \sum_{\ell=1}^{N_{\mathrm{beam}}} L_{\ell}$.

\begin{figure}[t!]
\centering
\includegraphics[width=1\columnwidth]{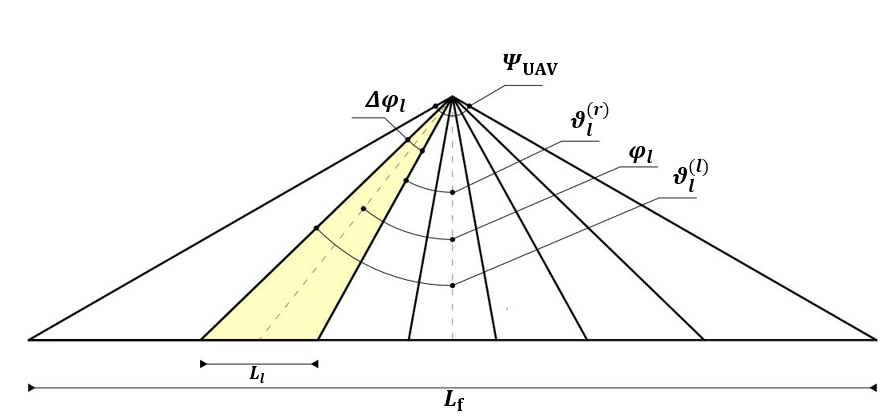}
\caption{Angles involved in the UAV beam footprint calculation.}
\label{fig:footprint}
\vspace{-0.5cm}
\end{figure}

As depicted in Figure \ref{fig:footprint}, the $\ell$th beam footprint can be computed geometrically as
\begin{equation}
    L_\ell = h_\mathrm{UAV} \left|\tan\left(\vartheta_\ell^{(l)}\right) - \tan \left(\vartheta_\ell^{(r)}\right)\right|,
\end{equation}
where $\vartheta_\ell^{(l)}$ and $\vartheta_\ell^{(r)}$ denote the left and right angle confining the $\ell$th beam, defined as
\begin{equation}
\begin{aligned}
\vartheta_\ell^{(l)} &= \varphi_\ell - \frac{\Delta\varphi_\ell}{2} \\
\vartheta_\ell^{(r)} &= \varphi_\ell + \frac{\Delta\varphi_\ell}{2},
\end{aligned}
\end{equation}
where $\varphi_\ell$ is the pointing direction of the $\ell$th beam, and $\Delta\varphi_\ell$ is the $\ell$th beamwidth, defined as
\begin{equation}
    \Delta\varphi_\ell = \frac{2}{N_\mathrm{UAV} \cos(\varphi_\ell)}.
\end{equation}

\vspace{-0.2cm}
\subsection{Average Access Probability Analysis}
\label{sec:covprob}

Herein, the objective is to analytically derive the average access probability, introduced in Sec. \ref{sec:intro},  for a \ac{CAV} seeking to establish communication with another \ac{CAV} but facing a blocked \ac{V2V} link, thereby exploiting relay through \ac{UAV}.

Let us consider a highway road segment of length $L_{\rm f}$ with $\rm K$ lanes. The length of the road segment depends on the height of the \ac{UAV} $h_\mathrm{UAV}$ and its covered field of view $\Psi_\mathrm{UAV}$, and it can be computed as
\vspace{-0.3cm}
\begin{equation}
    L_{\rm f} = 2 \, h_\mathrm{UAV} \, \cos\left(\frac{\Psi_\mathrm{UAV}}{2}\right).
\end{equation}

To determine the distribution of \acp{CAV} as a function of traffic density, we partition the highway road segment into spatial slots, each of size $L_{\rm v}$, with the capacity to accommodate a single \ac{CAV}. Consequently, $L_{\rm v}$ takes into consideration the physical size of the \ac{CAV} and allows for a minimum safety distance, assuming the scenario of maximum traffic congestion \cite{Kai}. The maximum number of \acp{CAV} in the segment covered by the \ac{UAV} is defined as $M^\mathrm{(max)} = \lfloor K L_{\rm f} / L_{\rm v} \rfloor$, while the maximum number of \acp{CAV} within the coverage area of the $\ell$th beam is expressed as $M_\ell^\mathrm{(max)} = \lfloor K L_\ell / L_{\rm v} \rfloor$. 
The probability that a spatial slot is occupied by a \ac{CAV} can be obtained assuming a Point Poisson distribution over a single slot length, as
\begin{equation}\label{occupationProb}
    P_\mathrm{o} = \lambda L_{\rm v} e^{-\lambda L_{\rm v}},
\end{equation}
where $\lambda$ is the vehicle density. 
It is evident that if $M_\ell^\mathrm{(max)} \leq N_\ell$, $\forall \ell$, the \ac{UAV} is capable of granting access to all incoming requests for relay services. Conversely, the average access probability is computed as 
\begin{equation}
   P^\mathrm{(ACC)} = \sum_{\ell=1}^{N_\mathrm{beam}} \frac{P^\mathrm{(ACC)}_\ell = \mathrm{Prob}\left(\overline{M}_\ell \leq N_\ell\right)}{N_\mathrm{beam}},
\end{equation}
where $\overline{M}_\ell$ denotes the number of valid relaying requests for the $\ell$th beam and $N_\ell$ is the maximum number of resources available for the $\ell$th beam, according to the RRA strategy in \eqref{fair} and \eqref{beambased}. 
Herein, we define a relaying request from the $i$th \ac{CAV} as a valid request if $\mathrm{SNR}_i \geq \gamma_\mathrm{th}$. The probability of having a valid request can be computed as
\begin{equation}
    P^\mathrm{(VR)}_\ell = \mathrm{Prob}\left(\mathrm{SNR}_\ell \geq \gamma_\mathrm{th}\right),
\end{equation}
where $\mathrm{SNR}_\ell$ is the average \ac{SNR} within the $\ell$th footprint, and $\gamma_\mathrm{th}$ is a threshold depending on the \ac{QoS} requirements. By the central limit theorem, the distribution of the average \ac{SNR} in dB can be approximated as $\mathrm{SNR}_\ell \sim \mathcal{N}(\mu_\ell, \sigma_\ell^2)$, where the mean value $\mu_\ell$ is computed using the path loss model in \eqref{PL} considering the barycenter of the $\ell$th footprint, and the variance is $\sigma_\ell^2 = \sigma_s^2 + \sigma_n^2$ \cite{LinsalataLoSMap, SM}. Hence, the probability of a  valid request is given by
\vspace{-0.2cm}
\begin{equation}\label{validReq}
    P^\mathrm{(VR)}_\ell = Q\left(\frac{\gamma_\mathrm{th} - \mu_\ell}{\sigma_\ell}\right),
\end{equation}
where $Q(\cdot)$ is the Q-function. 
\begin{figure*}[h!]
\begin{equation}\label{PACC}
    P^\mathrm{(ACC)}_\ell = \sum_{m=1}^{N_\ell} \binom{M_\ell^\mathrm{max}}{m} \, \left(P_\mathrm{o} \, P^\mathrm{(VR)}_\ell\right)^m \, \left(1 - P_\mathrm{o}\, P^\mathrm{(VR)}_\ell\right)^{M_\ell^\mathrm{max} - m}
\end{equation}
\hrulefill
\end{figure*}

Through the discretization of space within each footprint, we can compute the probability of access employing a combinatorial methodology. The rationale behind this approach is straightforward: with a total of $M_\ell^\mathrm{max}$ spatial slots available, the likelihood of the $m$th slot being occupied is determined by the computed probability $P_\mathrm{o}$ in Equation \eqref{occupationProb}, while the probability of a \ac{CAV} in the $m$th slot issuing a valid request is represented by $P^\mathrm{(VR)}_\ell$, as indicated in Equation \eqref{validReq}. Hence, the probability of access can be derived utilizing Equation \eqref{PACC}.

\begin{table}
\caption{Input Parameters}
\setlength{\tabcolsep}{3pt}
\begin{tabular}{|p{117pt}|p{35pt}|p{70pt}|}
\hline
Parameter & Notation & Value \\
\hline
CAV's antenna elements            & $N_{\rm UE}$ & 4 \\
\ac{UAV}'s antenna elements     & $N_{\rm UAV}$ & 8 \\
RF chains                       & $N_{\rm RF}$ & 4 \\
CAV's transmitting power          & $P_{i}$ & 23 [dBm] \\
Noise power                     & $\sigma_n^2$ & -101 [dBm] \\

Excess path loss offset      & $A$ & 84.64 [dB] \\
Path loss exponent     & $\alpha$ & 1.55 \\
Log-normal shadowing variance       & $\sigma_s^2$ & 4 \\

Number of lanes                 & $\rm K$ & 5 \\
UAV's Field of View             & $\Psi_{\rm UAV}$ & 120° \\
Vehicle's average length          & $L_{\rm v}$ & 5 [m] \\
Vehicle density          & $\lambda$ & 40, 80 [cars/km]  \\
SNR threshold          & $\gamma_\mathrm{th}$ & 5, 10 [dB] \\
Carrier frequency               & $f_o$ & 28 GHz \\
%Sub-carrier spacing & $\Delta f$ & 120 [kHz] \\
%Number of sub-carriers per RB & $N_{\text{sub}}$ & 12 \\
%Number of RBs per channel & $N_{\text{RB}}$ & 10 \\
Number of frequency
subchannels              & $N_{\text{ch}}$ & 2 \\
Slot duration                   & $T_\mathrm{slot}$ & $125$ [$\mu$s] \\
E2E max Delay                   & $\tau_\mathrm{E2E}$.  & $10$ [ms] - as in \cite{5GAA} \\
\hline
\end{tabular}
\label{tab1}
\end{table}
\vspace{-0.1cm}

\section{Numerical Results}
\label{sec:results}

This section presents the numerical results on the average access probability of a user as a function of both the \ac{UAV} altitude and the vehicle density and the performance comparison between the two \ac{RRA} techniques proposed. This comparison covers various scenarios, including different \ac{SNR} thresholds and varying traffic conditions. In this context, we evaluate and compare the number of vehicles with an \ac{SNR} above the threshold (connected vehicles) to the actual number of covered vehicles that have resources assigned (served vehicles) when both the \ac{RRA} methods are used.
The mathematical model results are validated via comparison with simulation results. The simulations were conducted in the MATLAB environment using parameters summarized in Table I. The simulated scenario faithfully replicates the description in Section \ref{sec:refscen}, with the sole distinction being the consideration of the actual \ac{SNR} between the drone and \ac{CAV}, as per Equation (4). In contrast, in the mathematical model, we assume vehicles under the same beam are characterized by the same \ac{SNR}, as outlined in Section \ref{sec:covprob}.
\begin{figure*}[!ht]
\centering
\begin{subfigure}{0.45\linewidth}
  \centering
  \includegraphics[width=0.8\linewidth]{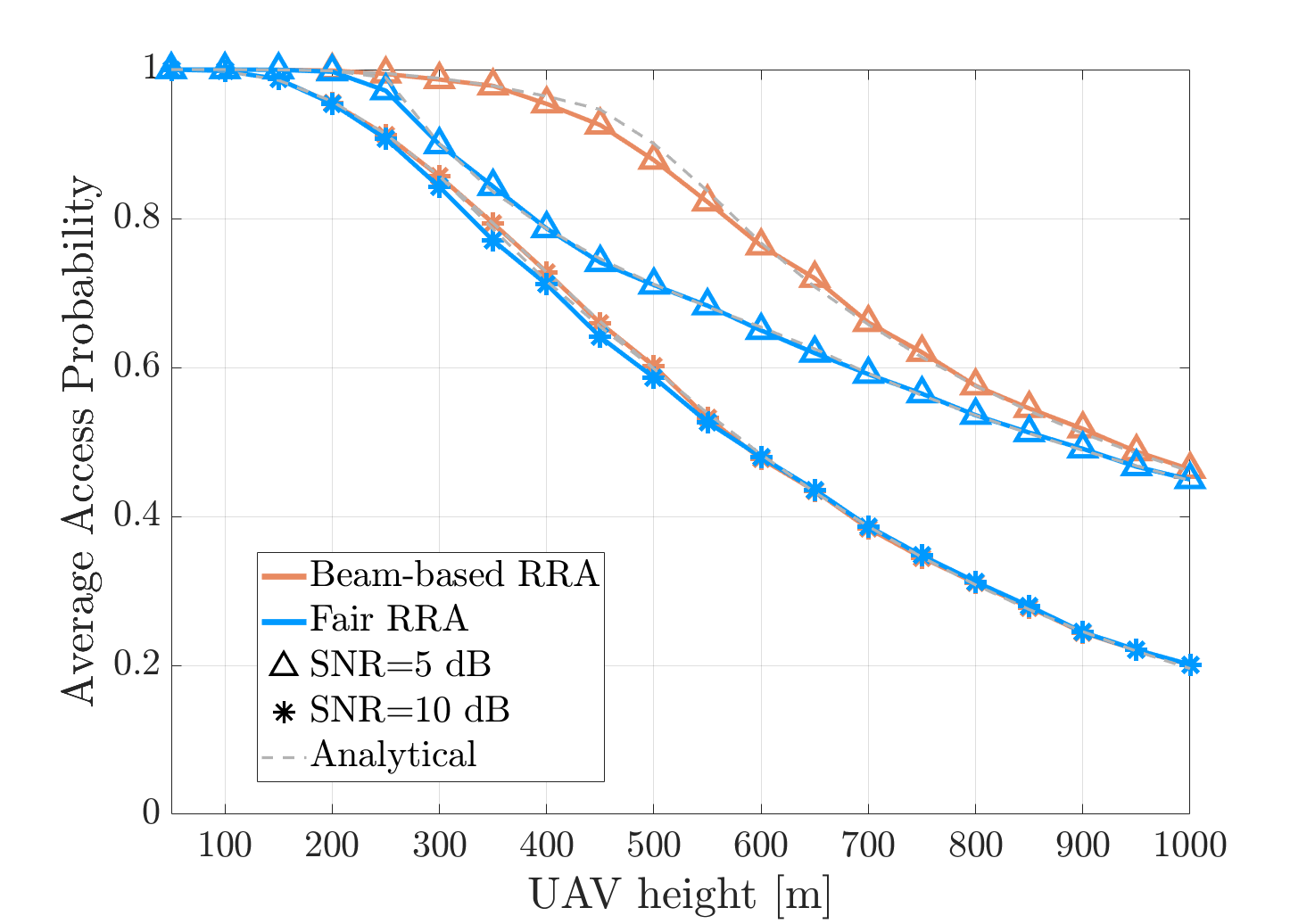}
  \caption{}
  \label{fig:res_1}
\end{subfigure}
\hspace{0.05\linewidth}
\begin{subfigure}{0.45\linewidth}
  \centering
  \includegraphics[width=0.8\linewidth]{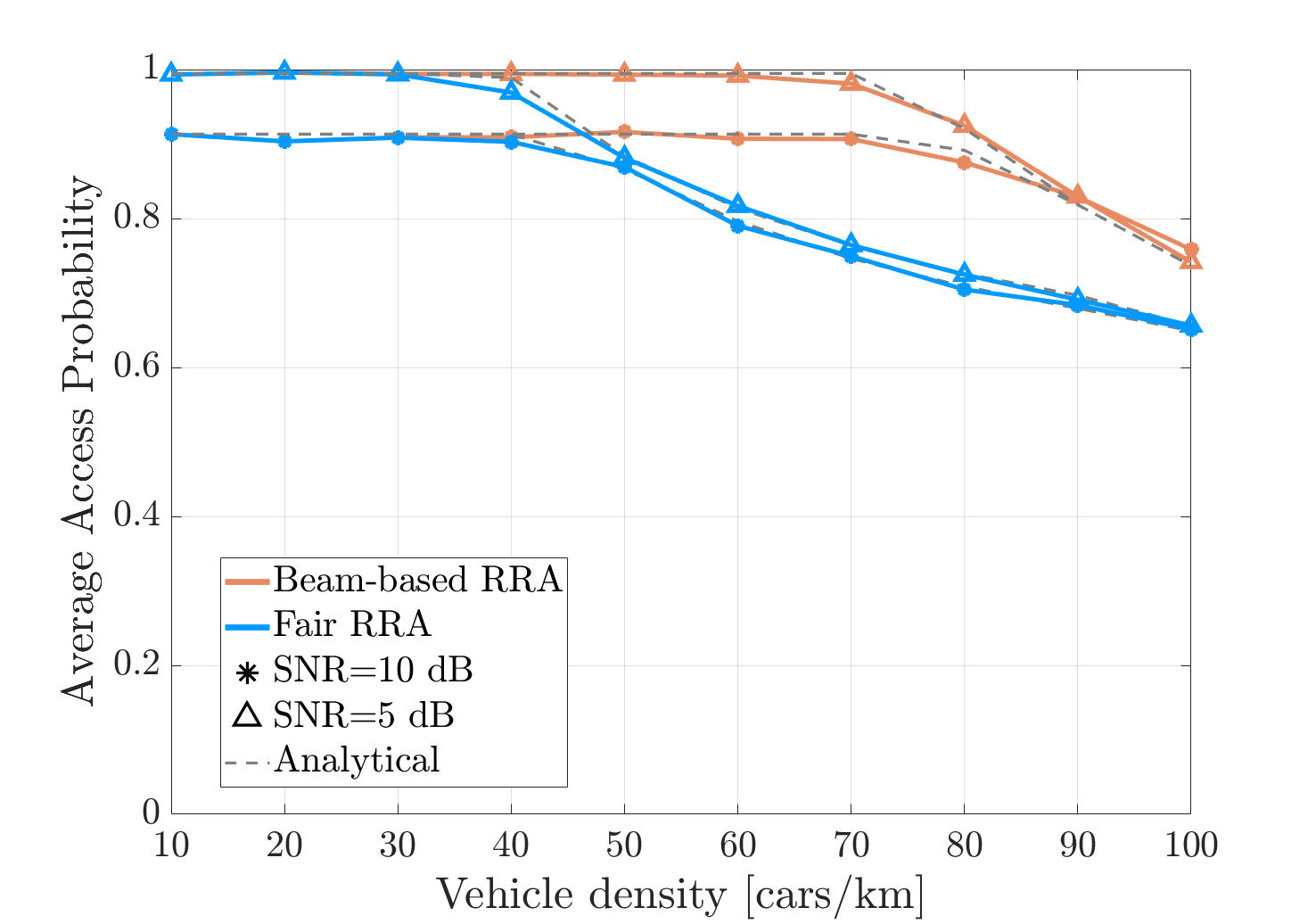}
  \caption{}
  \label{fig:res_2}
\end{subfigure}
\caption{Comparison of average access probability versus UAV height (a) and vehicle density (b) for different conditions.}
\vspace{-0.3cm}
\end{figure*}
\begin{figure*}[!ht]
\centering
\begin{subfigure}{0.45\linewidth}
  \centering
  \includegraphics[width=0.8\linewidth]{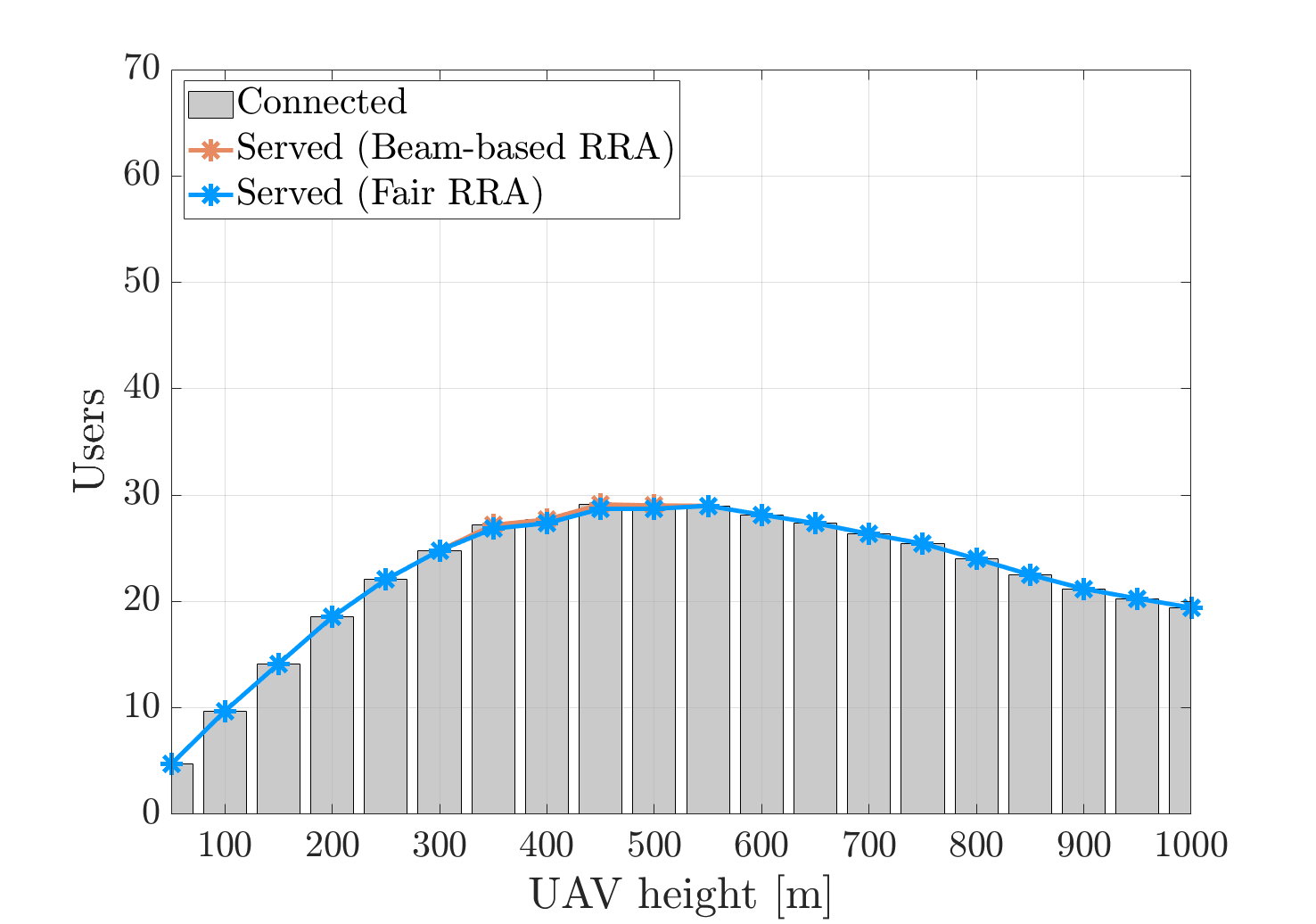}
  \caption{$\lambda=40$, SNR=10 dB.}
  \label{fig:res_3}
\end{subfigure}
\hspace{0.05\linewidth}
\begin{subfigure}{0.45\linewidth}
  \centering
  \includegraphics[width=0.8\linewidth]{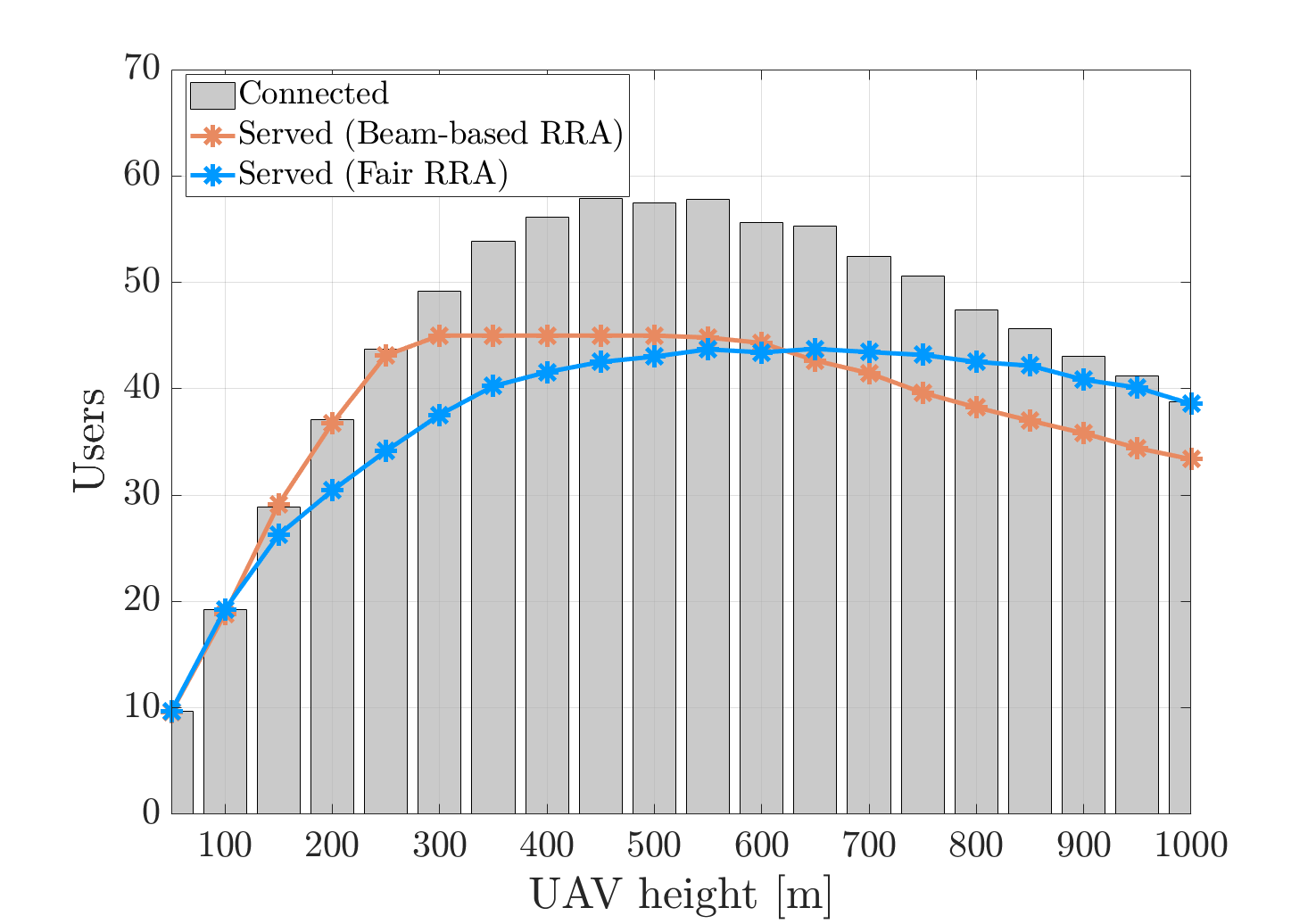}
  \caption{$\lambda=80$, SNR=10 dB.}
  \label{fig:res_4}
\end{subfigure}
\caption{Connected users versus served users with both RRAs. }
% \vspace{-0.3cm}
\end{figure*}

In Figure \ref{fig:res_1}, we report the average access probability as a function of the \ac{UAV} altitude. The curves were generated by maintaining a traffic density $\lambda$ constant at 40 cars/km and varying the \ac{SNR} threshold, $\gamma_{\rm th}$. In all the cases, the model's validity is confirmed by the near-perfect overlap of all four simulation curves with their respective theoretical counterparts (dashed curves). Looking at the two couples of curves in Figure \ref{fig:res_1}, it is clear how the gap between the performance of the two \ac{RRA} algorithms varies depending on $\gamma_{\rm th}$. Specifically, when the \ac{SNR} threshold is equal to 10 dB, both \acp{RRA} exhibit identical values of the average access probability. This suggests that, with such a high threshold, a significant portion of users is unable to access the network due to coverage limitations. Consequently, the number of users demanding resources becomes limited. As a result, both the beam-based \ac{RRA} and the fair \ac{RRA} effectively manage this reduced user population. When we increase the \ac{SNR} threshold to 5 dB, we can observe a significant improvement in performance when adopting the beam-based approach. In this case, more users will have coverage, and this is especially true as the drone's altitude increases because the beams footprints of the various beams will be larger and can accommodate more users. Hence, the beam-based \ac{RRA} stands out as the ideal algorithm to manage this heightened demand effectively. Concerning the overall decreasing trend observed, it is evident that as the drone's altitude increases, the radio link quality progressively deteriorates. Consequently, the average access probability, which represents the likelihood that vehicles are adequately covered and receive sufficient resources, decreases sharply due to adverse radio conditions. For an \ac{SNR} threshold of 10 dB, in order to meet the requirement of achieving a 99\% average access probability, the drone should fly at an altitude of 150 meters. If we reduce the \ac{SNR} threshold to 5 dB, then with the fair \ac{RRA}, the \ac{UAV} should operate at an altitude of 240 meters, whereas with the beam-based \ac{RRA}, it should be at 350 meters. From this information, if our goal is to provide \ac{V2X} service for a specific highway road segment, taking into account that the footprints of adjacent drones do not overlap, we can then derive the number of drones needed to effectively cover that particular stretch. 

The superiority of the beam-based \ac{RRA} becomes even more apparent when we plot the average access probability against varying vehicle density, as shown in Figure \ref{fig:res_2}. The two pairs of curves correspond to the two previously mentioned \ac{SNR} scenarios, 5 dB and 10 dB. Here, we set the drone's altitude at 250 meters. Consequently, with a 10 dB \ac{SNR} threshold, the average access probability never exceeds 91\%. However, with a 5 dB threshold, we achieve good performance, at least until the vehicle density surpasses 70 cars/km.

Figure \ref{fig:res_3} illustrates the comparison between the number of connected users (grey barplot) and the users effectively served controlling the \ac{UAV} altitude. The blue curve represents the fair \ac{RRA}, while the orange curve represents the beam-based \ac{RRA}. In this scenario, we maintain a constant \ac{SNR} of 10 dB and a $\lambda$ value of 40. This figure highlights a peak in connected users with increasing altitude. As the drone's altitude rises, its beams cover a larger area, accommodating more users. However, beyond a certain maximum altitude, the number of connected users declines due to deteriorating link quality, especially in the outermost beams. In this scenario, neither \ac{RRA} outperforms the other, for the reasons discussed earlier. 

To better appreciate the distinction between the two \ac{RRA} curves, we have doubled the value of lambda from 40 to 80 cars/km, thus moving to a higher \acp{CAV} density scenario, as in Figure \ref{fig:res_4}.
Here, we can observe two phenomena: firstly, the beam-based approach outperforms the fair \ac{RRA}. However, beyond a certain \ac{UAV}'s height, the situation reverses. At very high altitudes, the side beams, although populated by a large number of users, may struggle to establish connections. As a result, most users requiring resources are concentrated in the central beams. Since the fair \ac{RRA} allocates the same number of resources to all beams (with more resources assigned to central beams than the beam-based approach), it shows better performance in this specific context.

\section{Conclusions}
\label{sec:conclusions}
In this paper, by leveraging on beamforming-enabled \acp{UAV} as relay nodes, we have proposed a robust solution to ensure dependable data exchange, even in scenarios where direct \ac{V2V} communication faces hindrances.
The introduction of the novel performance metric, the average access probability, has provided a nuanced understanding of the satisfaction level among vehicles covered by a drone's beam during resource requests for uplink traffic demands. This metric serves as a crucial indicator for evaluating the efficacy of the proposed system.
Our contribution extends to the introduction of optimized \ac{RRA} algorithms tailored for \acp{UAV}. These algorithms consider several factors, including beamforming, resource availability, user distribution, and application latency requirements. The numerical results validate the proposed model, emphasizing the pivotal role of access probability in optimizing system parameters such as \ac{UAV} altitude and the number of drones.
This work, therefore, not only addresses the challenges inherent in \ac{V2V} communication but also presents a concrete framework for enhancing \ac{QoS} in \ac{V2X} sidelink communications. The findings of this study pave the way for further advancements in the integration of \acp{UAV} as relay nodes, contributing significantly to the realization of safer and more efficient connected vehicular networks.
\vspace{-0.2cm}

\section*{Acknowledgment}

This article was supported by the European Union under the Italian National Recovery and Resilience Plan (NRRP) of NextGenerationEU, partnership on “Telecommunications of the Future” (PE00000001 - program “RESTART”, Structural Project 6GWINET). Furthermore, it is based upon work from COST Action INTERACT, CA20120, supported by COST (European Cooperation in Science and Technology).
% \vspace{-0.2cm}
\bibliographystyle{IEEEtran}
\bibliography{IEEEabrv,StringDefinitions,v3}

\end{document}